\documentclass[11pt]{article}
\usepackage{moriond,epsfig,amssymb}

\def\Journal#1#2#3#4{{#1} {\bf #2}, #3 (#4)}

\def\NPB{{\em Nucl. Phys.} B}
\def\PLB{{\em Phys. Lett.}  B}
\def\PRL{\em Phys. Rev. Lett.}
\def\PRD{{\em Phys. Rev.} D}

\def\AA{{\em Astron.Astrophys.}}
\def\NAT{{\em Nature}}
\def\APJ{{\em Astrophys.J.}}

\def\be{\begin{equation}}
\def\ee{\end{equation}}
\def\bea{\begin{eqnarray}}
\def\eea{\end{eqnarray}}

\begin{document}
\vspace*{4cm}
\title{ANTI-MATTER IN COSMIC RAYS : BACKGROUNDS AND SIGNALS.}

\author{ Timur DELAHAYE }
\address{ LAPTH, Universit\'e de Savoie, CNRS,\\
BP110, F-74941 Annecy-le-Vieux Cedex, France}
\address{Dipartimento di Fisica Teorica, Universit\`a di Torino \\
Istituto Nazionale di Fisica Nucleare, via P. Giuria 1, I--10125 Torino, Italy}
\author{Pierre BRUN}
\address{CEA, Irfu, Service de Physique des Particules, Centre de Saclay, F-91191 Gif-sur-Yvette, France}
\author{ Fiorenza DONATO, Nicolao FORNENGO, Julien LAVALLE, Roberto LINEROS}
\address{Dipartimento di Fisica Teorica, Universit\`a di Torino \\
Istituto Nazionale di Fisica Nucleare, via P. Giuria 1, I--10125 Torino, Italy}
\author{ Richard TAILLET and Pierre SALATI }
\address{LAPTH, Universit\'e de Savoie, CNRS,\\
BP110, F-74941 Annecy-le-Vieux Cedex, France}
LAPTH-Conf-1321/09
\maketitle\abstracts{
Recent PAMELA and ATIC data seem to indicate an excess in positron cosmic rays above $\sim$ 10 GeV which might be due to galactic Dark Matter particle annihilation. However the background of this signal suffers many uncertainties that make our task difficult in constraining Dark Matter or any other astrophysical explanation for these recent surprising data. }

\section*{Introduction}
Recent cosmic ray measurements \cite{pamela_e,pamela_p,atic} have created a lot of excitation in the Dark Matter community.  Cosmic rays have been studied for many years and have given extremely satisfactory agreement between theory and observations for various species. However PAMELA \cite{pamela_e} and then ATIC \cite{atic} have changed this bright situation by confirming a discrepancy between theoretical expectations and observations for high energy ($\gtrsim$ 10 GeV) electrons and positrons that had been previously suspected thanks to HEAT\cite{1997ApJ...482L.191B} experiment. After stressing the various uncertainties that affect the description of the life of a cosmic ray, from creation to detection, we will discuss how difficult it is to find a correct explanation for all experimental data at once.\\
Even though very accurate methods \cite{galprop} have been developed, the estimation of the flux of positrons, electrons and anti-protons still suffers from many uncertainties. Most of cosmic rays (electrons, protons and heavier nuclei) are considered as primaries, namely particles accelerated by astrophysical objects of our galaxy (probably Super Nov\ae~ Remnants). However some species are believed to be secondaries which means that they are produced not by astrophysical objects but by other cosmic rays. The main process that produces positron, anti-proton but also Boron cosmic rays is believed to be the spallation of primary cosmic ray nuclei (mainly protons and $\alpha$ particles) on the Interstellar Medium (Hydrogen and Helium). In the case of anti-protons, one should even expect tertiary cosmic rays created by the spallation of secondaries. After production cosmic rays propagate in the galactic turbulent magnetic field and finally reach the Earth.\\
The idea of indirect detection is that, if Dark Matter exists and is coupled to the Standard Model sector, then the galactic halo should host annihilation (or decay) of Dark Matter particles hence producing a new primary component for cosmic rays. This process should produce as much matter as anti-matter but the background being much higher for matter. If any, the primary component due to Dark Matter has better chance to be found among anti-matter cosmic rays (at least with charge discriminating experiments).

\section{The various sources of uncertainties}
Though extremely appealing, indirect detection requires carefulness and disentangling a signal is possible only if the background is well understood. Here we will focus on the various facts that make our task difficult in evaluating the background.

\subsection{Cross sections}
Even though colliders give us plenty of data on proton-proton reactions at low energy, extrapolation to the energies relevant for cosmic rays are anything but trivial. The various available parametrizations may induce an uncertainty that may reach a factor of $3$ on the expected secondary positrons flux. Though a little less, anti-protons estimations are also affected by this uncertainty. Unfortunately this uncertainty do not just change the normalisation or the spectral index but even the shape of the spectrum which can be different from a power law.

\subsection{Proton flux}
\label{ssec:proton}
In order to correctly estimate the production rate of positron and anti-proton cosmic rays, one also needs to know the proton (and $\alpha$) flux everywhere in the galaxy and the matter density. However the only available data is of course the local value of the proton spectrum. Many experiments have measured this flux and most of them are consistent. However the radial distribution of proton sources and of Inter-Stellar Matter is less well known. Concerning positrons, this is of little importance indeed, because of energy losses, positrons are produced locally (see \ref{ssec:losses}) and are not sensitive to large scale radial variations neither of the proton flux nor of the matter distribution. For anti-protons the sensitivity to the sources radial distribution is a little larger.

\subsection{Energy losses}
\label{ssec:losses}
At the energy range we are interested in, energy losses concern only positrons. Some energy losses (e.g. Bremsstrahlung, ionisation of the Inter-Stellar Medium and adiabatic losses accompanying convection) are only relevant at energies lower than $\sim 10$ GeV where solar modulation dramatically affects the flux and make any comparison between data and prediction extremely dubious. But, the main energy losses, namely synchrotron radiation due to the steady part of the galactic magnetic field and inverse Compton scattering off of positrons on stellar, dust and CMB light, make positrons (and electrons) a very special species in the cosmic ray framework. Because we know this effect is important, positrons we detect at the Earth have to be created in the solar vicinity (around 80\% of the background at 10 GeV comes from less than 1 kpc). Measurements of galactic magnetic field and Inter-Stellar Radiation Field exist but both are affected by complex systematic effects. Even though we limit ourselves to the local region where positrons are created, the uncertainty on the typical energy loss time scale is still of order $\sim 3$. This translates into very small variation of the shape of the expected flux but also in a change of the normalisation by a factor  $\sim \sqrt{3}$. This uncertainty also drastically affects how far the sources of the positrons we detect on Earth are. As it will be shown in \ref{sec:signal}, if interested in interpreting the PAMELA result in term of a point-like source, one cannot change the normalisation of the flux without affecting the number and distance of sources.

\subsection{Propagation}
\begin{figure}
\begin{center}
\psfig{figure=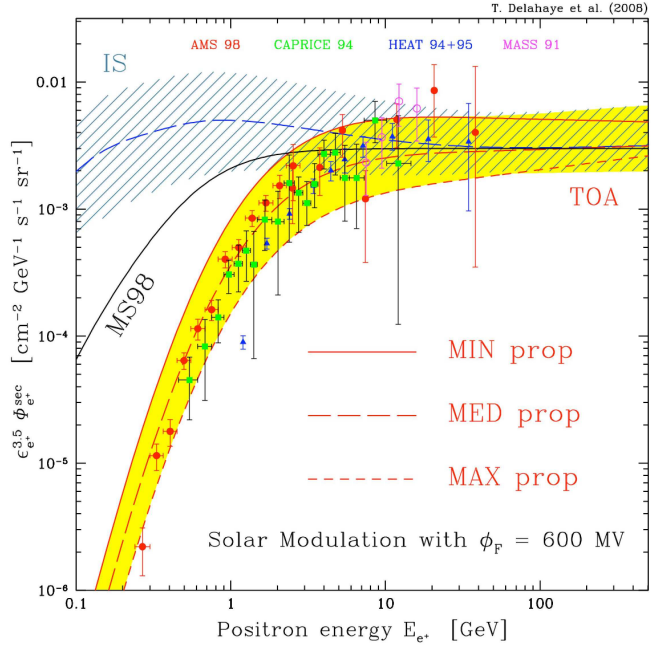,height=9cm}
\end{center}
\caption{Secondary positron flux as a function of the positron energy. The blue hatched 
band corresponds to the CR propagation uncertainty on the Inter-Stellar prediction 
whereas the yellow strip refers to Top Of the Atmosphere fluxes.
The long--dashed curves feature our reference model with the Kamae$^{14}$
parameterization of nuclear cross sections, the Shikaze$^{19}$ injection
proton and helium spectra and the MED set of propagation parameters.
The MIN, MED and MAX propagation parameters displayed are from Maurin \textit{et alii}$^{16}$ .
Data are taken from the Caprice$^6$, HEAT$^5$, AMS$^{3,4}$ and MASS$^{13}$ experiments.}
\label{fig:uncertainty}
\end{figure}

Charged particles do not travel easily in the galaxy : they scatter off on the inhomogeneities of the galactic magnetic field and are re-accelerated by them, they convect under the pression of the galactic wind, they loose energy, some of them decay and they interact with the Interstellar Medium through spallation processes. All these processes are summed up in the diffusion equation \ref{eq:dif} that has to be solved with the proper boundary conditions.
\begin{equation}
\mathbf{\nabla} \! \cdot
\left\{
- \, K_0 \, \epsilon^{\delta} \, \mathbf{\nabla} N \, + \,
\mathbf{V}_{C}(z) N
\right\} \; 
+  {\displaystyle \frac{\partial}{\partial \epsilon}}
\left\{
b^{\rm loss}(\epsilon,z) \, N \, - \,
K_{\epsilon \epsilon}(z) \, \frac{\partial N}{\partial \epsilon}
\right\}  =  q_{e^{+}}(\mathbf{x},\epsilon) 
\label{eq:dif}
\end{equation}
 Observations of other galaxies suggest that cosmic rays are diffusing in a cylindric slab, the height of which seems to vary from one galaxy to another. As soon as a cosmic ray reaches an edge of the diffusion zone, it is expected to leave the zone and to never return. All these processes are not very well constrained neither theoretically nor observationally. However the ratio of secondary over primary cosmic rays depends almost only on propagation. Using Boron/Carbon data, Maurin \textit{et alii} \cite{B/C} have constrained the values of the parameters of the propagation equation \ref{eq:dif}. However, even under these constraints, the compatible parameter space is still quite extended and sizing the underlying uncertainty requires to scan the complete parameter space. This is why one needs a fast method to compute cosmic ray fluxes, which is allowed by our method. Depending on the energy we are interested in, the parameter set that maximizes (or minimizes) the flux is not always the same. This is why it is not enough to look at the envelope of fig. \ref{fig:uncertainty} to estimate uncertainties due to propagation and analysing data really requires a full scan of the parameter space.

\subsection{Electron flux}
At the moment, the PAMELA collaboration has not published absolute positron flux yet but the positron fraction (namely the flux of positron divided by the flux of positron plus the flux of electrons). There are $\sim 10$ times more electrons than positrons in cosmic rays, this is why any small uncertainty on electron flux has a dramatic effect on positron fraction. Though it seems quite improbable that PAMELA result can be explained by standard secondary positrons only, interpretation of the excess cannot be done without knowing its range and shape which clearly relies on the electrons. During this conference the PAMELA collaboration has presented a preliminary result concerning the spectral index of electron flux. It will considerably help future works and clarify the situation. However without the normalisation and the systematic errors, a lot of uncertainties will remain.

\subsection{Solar Modulation}
Because of its magnetic and coronal activities, the Sun perturbs low energy ($\lesssim $10 GeV) cosmic rays. The interaction between solar wind, heliosphere and cosmic ray fluxes is called solar modulation and is not very well understood yet. PAMELA results seem to indicate that the simplest model for this solar modulation is not correct and that in fact, cosmic ray charge and mass but also solar polarity may have a role to play here. PAMELA is the first apparatus which measures cosmic rays from space over such a long time. Therefore, unlike balloons which only give us a snapshot at one precise moment, PAMELA is able to look at the evolution of cosmic ray fluxes with solar activity. Time-dependent data are not available yet but let us hope that they will be soon. Indeed, as long as we are not able to model solar modulation properly, all the data that are affected by this phenomenon (namely the one below $\sim 10$GeV) are extremely difficult to analyse.

\section{Interpreting the signal}
\label{sec:signal}
\subsection{Degeneracy}
\begin{figure}
\begin{center}
\psfig{figure=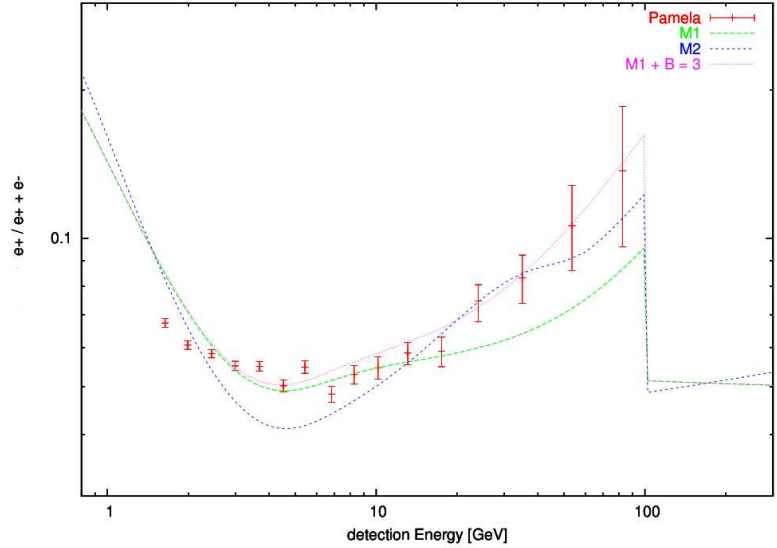,height=9cm,angle=0}
\end{center}
\caption{By changing the propagation parameters and the energy loss time scale for both primaries and secondaries, the same model of Dark Matter (here $\chi + \chi \rightarrow e^+ + e^-$ with $m_\chi = 100$GeV.c$^{-2}$) leads to different conclusions.}
\label{fig:exemple}
\end{figure}

There are many attempts to explain the PAMELA and ATIC features in the recent literature, most of them with pulsars or Dark Matter models. However to properly test any model, it is of utmost importance to use the same propagation parameters for both secondary and primary cosmic rays. Figure \ref{fig:exemple} shows why it is not licit to change the normalisation of the background and then fit the remaining of PAMELA data with one's favored Dark Matter model and any propagation model. In this plot we have considered an electron spectrum in agreement with the AMS data (which is the best one can do without more data from PAMELA), and a Dark Matter particle of 100 GeV.c$^{-2}$ annihilating only into electron/positron pairs and following the density computed by Moore et alii \cite{moore}. $M1$ and $M2$ correspond to two different propagation parameter sets with a energy loss typical time that have been chosen (in the range allowed by observations) to give approximatively similar secondary backgrounds. However when the primary cosmic rays are added, the results are quite different. In one case ($M2$), we find that an annihilation cross section of $<\sigma v> = 3.\times 10 ^{-26}$cm$^3$.s$^{-1}$ is enough to roughly agree with PAMELA data whereas on the other hand, with the second propagation model, ($M1$) we need a boost factor of $3$. For heavier Dark Matter particles or for other annihilation channels the discrepancy can be larger than one order of magnitude.
\\
The reason for this result is that the propagation parameters do not have the same importance for secondaries and primaries. Indeed, secondary positrons are mainly affected by the diffusion coefficients $K_0$ and $\delta$ of eq. (\ref{eq:dif}). But, for Dark Matter, the most important coefficient is the size of the diffusive halo $L$, which determines what fraction of the Dark Matter halo contributes to the signal. Moreover, the energy loss time scale, does not only change (at first approximation) the normalisation of the background, it also sizes the maximum distance from which a punctual source can participate to the flux at a given energy. Hence it limits the number of pulsars or Dark Matter clumps one is allowed to consider trying to explain the PAMELA data.
\subsection{Multi-channel analysis}
It is clear now that positron data are not enough to solve the PAMELA puzzle. Some uncertainties will diminish as soon as PAMELA will publish new data. Indeed absolute proton flux will alleviate the issues stressed in \ref{eq:dif}, Boron to Carbon ratio will give better constraints on propagation parameters, absolute electron flux will enable more serious analysis and temporal variation may help explaining solar modulation. However some degeneracy will remain. This is why it will be necessary to confront any model that tries to explain the positron excess to other data.
\\
Many species are under study. Neutrinos are not very constraining now (see e.g. Peter~\cite{neutrinos}) both because of theoretical uncertainty and experimental limits. Gamma rays also suffer from important theoretical uncertainty, mainly because of our poor knowledge of the Diffuse Emission and lack of observation. However, thanks to the new Fermi experiment, this will change very soon. Not only will this constrain luminosity of Dark Matter inhomogeneities (clumps) but also it will give better knowledge of nearby pulsars, constraining their number, distance and maybe amount of enery that can go to primary positrons and electrons. 
\\
Most important are the anti-protons : PAMELA has also published \cite{pamela_p} an anti-protons to protons ratio that is in very good agreement with theoretical expectations and leaves very little room for any excess bellow 100 GeV. This is probably the most challenging constraint in explaining the PAMELA data in terms of Dark Matter or alternative Cosmic Rays models. Indeed most common scenarios predict that Dark Matter is coupled to hadrons and therefore its annihilation should also give anti-protons. This means that, either Dark Matter is not much coupled to hadrons or it is extremely heavy ($\gtrsim$ few TeV.c$^{-2}$) \cite{cirelli,Donato:2008jk}.

\subsection{ATIC}
The balloon borne experiment ATIC \cite{atic} has claimed to detect an excess around 600 GeV in the measured positron plus electron spectrum. Interpretation of this signal is even more controversial than the PAMELA excess. Indeed, astrophysical processes that can produce high energy electrons are much more numerous than for positrons, therefore it would be interesting to know whether or not positrons are involved in this feature. At these energies it is currently impossible to make a charge dependent detection. However, very soon, the Fermi experiment should publish an electron plus positron flux as well and this should decrease statistical errors. More interesting would be to have one or two more points from PAMELA. At 100 GeV only 15\% of leptonic cosmic rays are positrons. If ATIC result is due to Dark Matter or pulsars, then half of its signal should be made of positrons,  hence a positron fraction that should reach about 0.5 around 200 GeV. If PAMELA cannot reveal this, then AMS02, should give the answer.

\section*{Conclusions}
Present situation is far from clear in the cosmic ray landscape. The need for new data from PAMELA, ATIC and future experiments is obvious if we want to unshade the PAMELA puzzle. It is clear that these recent data are proof for either new physics or new astrophysics. However answering this question requires to be extremely cautious both dealing with backgrounds and signals.

\section*{Acknowledgments}
T.D. acknowledges financial support from the International Doctorate on AstroParticle Physics (IDAPP) and from the Explora'Doc exchange program of the Rh\^one-Alpes region.

\end{document}